\documentclass[journal, 10pt]{IEEEtran}

\usepackage[utf8]{inputenc}

\usepackage{siunitx}
\DeclareSIUnit\us{\micro\s}
\usepackage{booktabs}
\usepackage{xspace}
\usepackage{multirow}
\ifdefined\LINENUMBERS
\usepackage[switch]{lineno}
\fi
\usepackage{hyperref}
\usepackage[capitalise]{cleveref}
\Crefname{figure}{Fig.}{Figs.}
\Crefname{equation}{Equation}{Equation}
\crefname{equation}{}{}
\usepackage{todonotes}

\usepackage{cite}

\newcommand{\vulom}{VULOM4b\xspace}
\newcommand{\TRLOII}{TRLO II}
\newcommand{\LT}{\ensuremath{\mathcal{L}}}
\newcommand{\DT}{\ensuremath{\mathcal{D}}}
\newcommand{\BT}{\ensuremath{\mathcal{B}}}
\newcommand{\mLT}{\LT}

\newcommand{\fr}{\ensuremath{f_{\mathrm{r}}}}
\newcommand{\fa}{\ensuremath{f_{\mathrm{a}}}}
\newcommand{\fmax}{\ensuremath{f_{\mathrm{a,max}}}}
\newcommand{\tg}{\ensuremath{t_{\mathrm{g}}}}
\newcommand{\tc}{\ensuremath{t_{\mathrm{c}}}}
\newcommand{\tp}{\ensuremath{t_{\mathrm{p}}}}
\newcommand{\too}{\ensuremath{t_{\mathrm{o}}}}
\newcommand{\td}{\ensuremath{t_{\mathrm{d}}}}
\newcommand{\ts}{\ensuremath{t_{\mathrm{s}}}}
\newcommand{\tn}{\ensuremath{t_{\mathrm{n}}}}
\newcommand{\nb}{\ensuremath{n_{\mathrm{b}}}}
\newcommand{\ns}{\ensuremath{n_{\mathrm{s}}}}

\ifdefined\LINENUMBERS
\newcommand{\revlabel}[1]{\label{#1}\linelabel{#1}}
\else
\newcommand{\revlabel}[1]{\label{#1}}
\fi
\newcommand{\revlabelnolineno}[1]{\label{#1}}

\begin{document}
\bstctlcite{IEEEexample:BSTcontrol}
\title{VME Readout At and Below the Conversion Time Limit}

\author{
  M.~Munch, J.~H.~Jensen, B.~L\"oher, H.~T\"ornqvist, and H.~T.~Johansson%
  \thanks{%
    M.~Munch and J.~H.~Jensen are with the Department of Physics and Astronomy, Aarhus
    University, 8000 Aarhus C, Denmark.
  }%
  \thanks{%
    B.~L\"oher and H.~T\"ornqvist are with the Institut f\"ur Kernphysik, Technische
    Universit\"at Darmstadt, 64289 Darmstadt, Germany, and the GSI Helmholtzzentrum f\"ur
    Schwerionenforschung GmbH, 64291 Darmstadt, Germany.
  }%
  \thanks{%
    H.~T.~Johansson is with the Department of Physics, Chalmers University of Technology,
    SE-412 96 G{\"o}teborg, Sweden.
  }%
  \thanks{%
    The work of M. Munch was supported by the European Research Council under ERC
    starting grant LOBENA, No. 307447.
  }%
  \thanks{%
    The work of J.~H.~Jensen was supported by the Danish Council for Independent Research -- Natural Sciences under grant DFF -- 4181-00218.
  }%
  \thanks{%
    B.~L\"oher and H.~T\"ornqvist were supported by the GSI-TU Darmstadt cooperation agreement.%
  }%
  \thanks{%
    The work of H. T. Johansson was supported by the Swedish Research Council under grant
    822-2014-6644 and the Lars Hierta Memorial Foundation.
  }%
}

\IEEEpubid{0000--0000/00\$00.00}

\maketitle

\ifdefined\LINENUMBERS
\linenumbers
\fi

\begin{abstract}
    The achievable acquisition rates of modern triggered nuclear physics
  experiments are heavily dependent on the readout software,
  in addition to the limits given by the utilized hardware.
  This paper presents an asynchronous readout scheme that significantly improves the livetime of an otherwise synchronous triggered VME-based data acquisition system.
  A detailed performance analysis of this and other readout schemes, in terms of the basic data transfer operations, is described.
  The performance of the newly developed scheme as well as synchronous schemes on two systems has been measured.
  The measurements show excellent agreement with the detailed description.
  For the second system, which previously used a synchronous readout,
  the deadtime ratio is at a 20~kHz trigger request frequency reduced by 30~\%
  compared to the nearest contender, allowing 10~\% more events to be recorded
  in the same time.
  The interaction between the network and readout tasks for single-core processors is also investigated.
  A livetime ratio loss of a few percent can be observed, depending on the size of the data chunks given to the operating system kernel for network transfer.
  With appropriately chosen chunk size, the effect can be mitigated.

\end{abstract}


\IEEEpeerreviewmaketitle

\section{Introduction}

\IEEEPARstart{M}{ost} modern nuclear physics experiments %
have %
detectors, front-end electronics,
computer control, %
and network. The role of the front-end electronics is to digitize the detector signals
such that they can be analyzed.
Modular front-end electronics are typically housed in crates such as NIM, FASTBUS, CAMAC or VME,
where
the latter three also contain a bus on the crate backplane. A typical crate configuration consists of a group of
front-end modules together with a controlling single board computer (SBC).
The task of the SBC is to
acquire data from the front-end modules and transfer it over the network to permanent
storage and online analysis. The speed, overhead, and serialization effects
of this transfer naturally limit the maximum achievable acquisition rate, characterized by the rate of accepted triggers.

This article will focus on VME-based \cite{VITA1994} readout systems.  The purpose is not to
introduce new electronics. Instead, we aim at better utilizing the commercially available
modules by generally introducing an extreme asynchronous multi-event readout scheme called \emph{shadow
readout}.  The main speedup is achieved by almost completely decoupling the readout from the
conversion sequence, thereby significantly lowering the readout overhead associated with each
accepted trigger.
In addition, the coincidence information leading to each trigger is recorded, thus not sacrificing event selection flexibility for speed.
It is also shown that the remaining %
system deadtime can be accurately described based on the timing of basic
data transfer operations.

The article is structured in the following way: First, the existing solutions are reviewed, and
the necessary concepts in
order to model the deadtime and efficiency of the different readout modes
are introduced. This is followed by a brief discussion of various
modes of VME access. We then describe our implementation and the caveats that arise from
having a single-core SBC. Finally, we benchmark the improved readout scheme versus the
other strategies on two different systems.

\IEEEpubidadjcol

\section{Existing solutions}
\label{sec:existing-solutions}

With CAMAC systems, LeCroy introduced fast readout using the FERA bus %
in the early 80s,
which transfers data at 10 Mword/s (16-bit) \cite{lecroy4003b}.
With conversion times, at that time,
usually around \SI{3}{\us}/channel or \SI{25}{\us} for an
8-channel module, a crate with 50 words of data per event (after
zero-suppression) would be read out in \SI{5}{\us}.  Thus the
readout overhead was much smaller than the conversion time.

FASTBUS modules \cite{Larsen1982}, even though known for their high channel density and
thus long total module conversion times, usually had multi-event
buffers, allowing readout of previous events while converting
new ones, completely hiding the readout.

The current state of affairs for VME based readout systems is that nearly all front-end modules
are capable of storing multiple converted events in an internal memory buffer. 
The Daresbury MIDAS \cite{daresbury} and BARC-TIFR LAMPS \cite{LAMPS} utilize these buffers fully to
decouple readout from conversion.
In this scheme, the module buffers are continuously emptied by the
crate controller. Only if a module becomes full, will it assert a long busy; halting the
acquisition until the SBC empties a part of the module buffer.
This mode, however, is only available for the Silena S9418 front-end modules in MIDAS \cite{SAC}, and
the LAMPS is hampered by large overhead times in the associated CAEN VME controller \cite{Ramachandran_priv_may2018}.
\begin{figure}[!t]
  \centering
  \includegraphics[width=0.95\columnwidth]{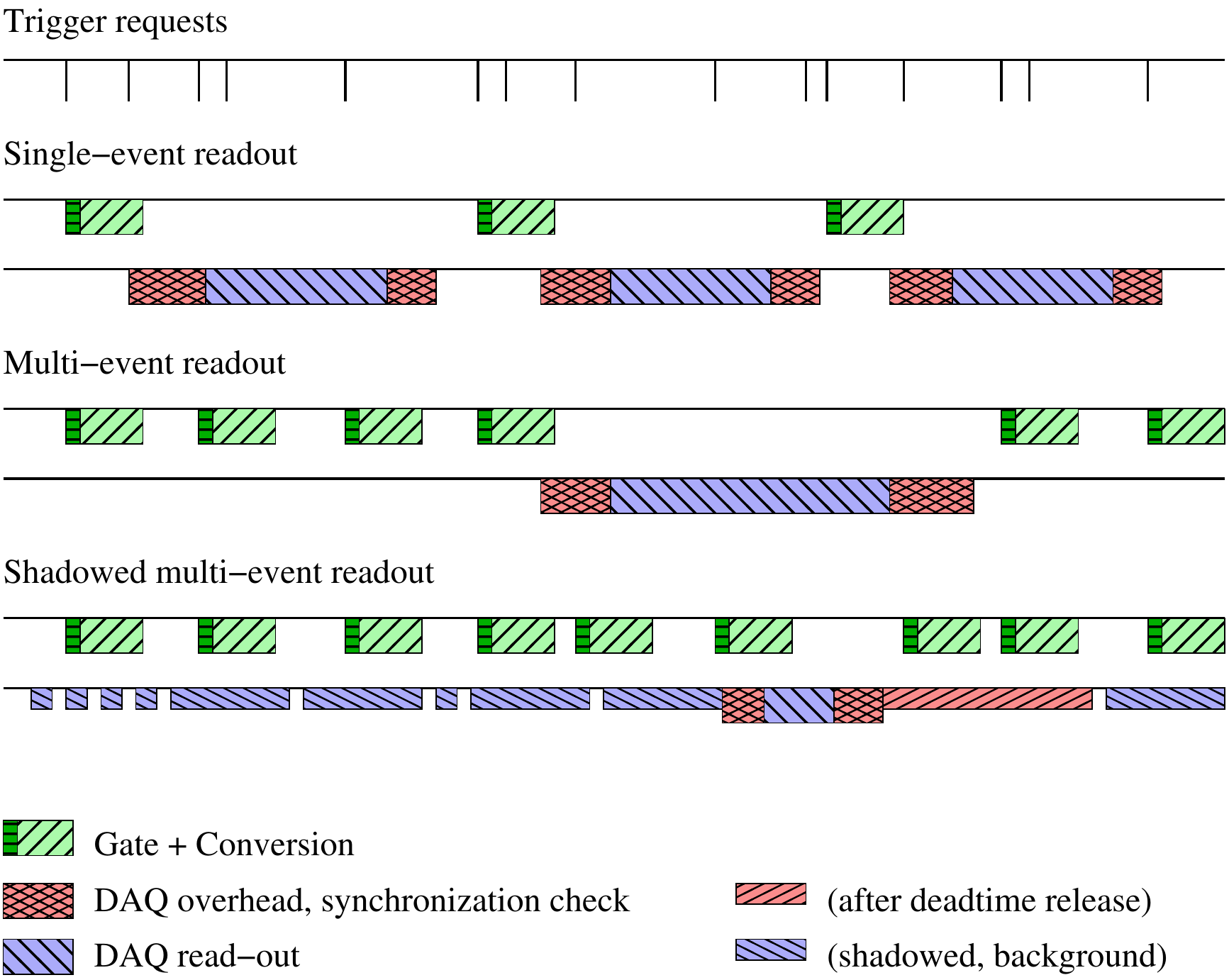}
  \caption{Single, multi-event, and shadow multi-event readout.
    Multi-event readout allows event data to accumulate in the module
    buffers before readout, amortizing the overhead over several
    events.  Shadow readout continuously empties the modules during
    conversions, thus reducing the work during global deadtime.}
  \label{fig:modes}
\end{figure}

However,
the above systems are exceptions, and
the multi-event buffers are often not utilized to their full extent, or not at all.
\Cref{fig:modes} illustrates
the three principal modes in which a readout system can be operated when front-end modules have
buffers. In the simplest case, the front-end acquires an event and waits for the readout to
transfer it. This is called single-event mode. Typically, the data transfer will
take significantly longer than the conversion time due to various overheads. These overheads
can be partially amortized by filling the front-end buffers before performing a
readout.
This is called multi-event mode.
However, the number of events that can be acquired in one go is limited to the
shallowest buffer. To the best of our knowledge, common general-purpose nuclear
physics data acquisition systems, such as MIDAS (PSI/TRIUMF) \cite{ritt1997midas}, MBS (GSI)
\cite{Essel2000}, and RIBFDAQ (RIKEN) \cite{Baba2010}, use versions of these schemes, where the SBC must interact with the trigger logic for every readout round.

In order to minimize the transfer times, when running in single event mode, the \revlabel{revMOCO}Mountable
Controller (MOCO) \cite{MOCO} was recently developed at RIKEN. This innovative design is an
adapter board with an FPGA and a USB interface %
which is installed between a front-end module and the VME crate backplane. With MOCO installed, the data
lines of the front-end module are not connected to the backplane. Instead, the FPGA communicates
directly with the front-end module. The FPGA can then transfer the data to the controlling computer
via USB. The two main benefits of this solution are cost and parallelization of the readout of
multiple modules.

In this article, we will show that a continuous readout mode,
operating beyond the depth of the multi-event buffers,
can be applied generally, without introducing new electronics.

\markboth{VME Readout At and Below the Conversion Time Limit}%
{VME Readout At and Below the Conversion Time Limit}
\section{Deadtime modeling} %
\label{sec:concept-modelling}

In this section, we develop a framework to describe the deadtime of a
system. This model will be used to analyze the systems in \cref{sec:benchmark}.

Assume that the front-end electronics and readout system is provided with a stream of
Poisson-distributed trigger requests, with an average frequency $\fr$,
of which it can accept some frequency $\fa$.  The livetime ratio is the
fraction of accepted trigger requests:
\begin{equation}
  \label{eq:lt}
  \LT = \frac{\fa}{\fr} = \frac{1}{1 + \fr\, \Delta t}. 
\end{equation}
$\Delta t$ is the amount of time per event that the system is blocked. The last equality
assumes the deadtime to be non-extending, and the formula is derived
in \cite{Knoll}. The livetime ratio is related to the deadtime ratio via 
$\DT = 1 - \LT = \fa\, \Delta t$.

We carefully differentiate between a system in deadtime
and the deadtime ratio, \DT{}.  A system in deadtime rejects events while
the deadtime ratio is the fraction of rejected events.  The same
differentiation is made between the livetime and the livetime ratio, \LT.

It is often necessary to consider a distinct contribution to the total
system deadtime: busy.
Busy signals come from front-end modules, and is also autonomously
released by those.  It is generally asserted during gate and
conversion, and when the data buffer of a module is full.
The more regular deadtime is initiated by the trigger logic, and
removed by the SBC readout software.
As far as trigger acceptance is concerned, both have the same effect:
inhibit triggers.

While running, the system must perform the following tasks (with associated processing times):
\begin{enumerate}
\item apply the gate of each accepted event, $\tg$;
\item convert the event, $\tc$;
\item poll for data, $\tp$ (per poll);
\item readout overhead (e.g.\ determine data amount), $\too$;
\item read data, $\td$;
\item check module synchronicity, $\ts$;
\item transfer data over the network, $\tn$.
\end{enumerate}
The two first are handled by the front-end electronics, the other by the SBC.
Some of the tasks happen during
either live- or deadtime, depending on the design of the system.
Thus the minimal
deadtime ratio is given by the fraction of time modules are busy converting data:
\begin{equation}
  \label{eq:busy}
  \BT = \fa (\tg + \tc).
\end{equation}

If the system is designed such that it reads one single event at a time, and will not accept new
events while reading data, then the readout time is $(\tp + \too + \td + \ts)$, since the system
must poll for data, determine the amount of data, read the data and check for synchronicity.
Since the likelihood that a random trigger request occurs during deadtime is simply the total
fraction of time the system will not accept triggers, the deadtime ratio is
\begin{equation}
  \label{eq:se}
  \DT = \BT + \fa (\tp + \too + \td + \ts).
\end{equation}

However, this scheme suffers the overhead of $\tp + \too + \ts$
for every event. If the front-end electronic modules have buffers of a certain depth $\nb$,
then the cost of polling, general readout overhead, and synchronization can be amortized:
\begin{equation}
  \label{eq:me}
  \DT = \BT + \fa\left(\td + \frac{\too + \tp + \ts}{\nb}\right).
\end{equation}
This mode of operation is commonly referred to as multi-event mode with the former called
single-event mode. 
In this case, $\td$ can often also be reduced as the larger amounts of
data per transfer may profit from faster transfer modes.

If, however, the system can simultaneously accept new events and transfer data, then \DT{}
may be %
further reduced:
\begin{equation}
  \label{eq:shadow}
  \DT = \BT + \fa\frac{\alpha \td + \too + \tp + \ts}{\ns}.
\end{equation}
$\ns$ is the number of events accepted before synchronicity is checked and 
$\alpha$ is the average number of events remaining to read out during deadtime. The fraction of events read during deadtime is thus $\alpha/n_{s}$. $\alpha \ge 0$ but will typically only be a few events. %
It should be
noted that $\nb$ for many older modules is typically 32 or less, while newer modules might store
a few hundred events. However, $\ns$ can in principle be made 
arbitrarily large, and thus
the deadtime can be reduced to the busy time, $\DT \approx \BT$.

\begin{figure}[tb]
  \centering
  \includegraphics[width=\columnwidth]{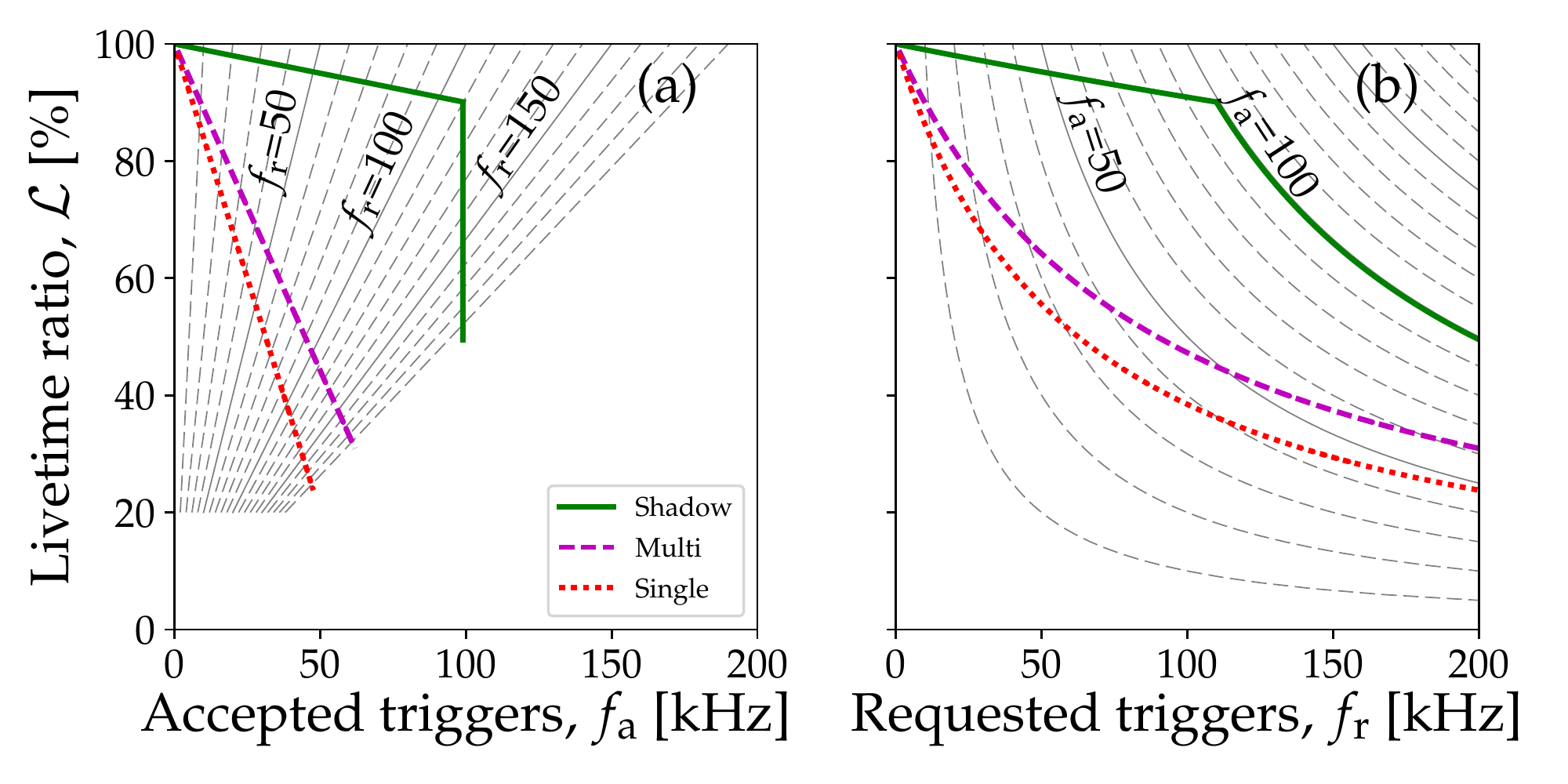}
  \caption{Schematic performance of shadow, multi-event, and single-event 
    modes, with \revlabelnolineno{revgenericlt}$\tg+\tc=\SI{1}{\us}$,
    $\tp+\too=\SI{3}{\us}$, $\td=\SI{10}{\us}$, $\ts=\SI{2}{\us}$,
    $\nb = 32$, and $\ns = 1000$.
    Both figures show the same systems, (a) as
	a function of $\fa$ and (b) as a function of $\fr$.  The 
	background lines show constant $\fr$ and $\fa$, respectively.
  }
  \label{fig:genericlt}
\end{figure}

The above has not considered the available data transfer bandwidth,
which may also limit the maximum event rate that can be accepted:
\begin{equation}
  \label{eq:fmax}
  \fmax = \frac{1}{(\td + \frac{\too + \tp}{\nb} + \frac{\ts}{\ns})}.
\end{equation}
In practice, the achievable frequency will be slightly lower due to various other overheads
such as %
network transfer. %
Above this request frequency, the
livetime ratio just deteriorates while no more events are accepted and
is simply described:
\begin{equation}
  \label{eq:saturated}
  \LT = \frac{\fmax}{\fr}.
\end{equation}
The livetime of the system is thus given by whichever of \cref{eq:lt} or
(\ref{eq:saturated}) is smaller:
\begin{equation}
  \label{eq:multi}
  \mLT = \min\left(\frac{1}{1 + \fr \Delta t}, \frac{\fmax}{\fr}\right).
\end{equation}
For shadow readout, $\Delta t$ is expected to be approximately equal to the gate and conversion time $\tg + \tc$.
The livetime characteristics of the different readout schemes are
illustrated in \cref{fig:genericlt}, where also
the hard limit nature of $\fmax$ is clearly seen.

\revlabel{revsmooth}%
In practice, the transition between the two domains is not sharp,
due to interplay among the different saturating mechanisms,
which is not investigated further.
However, a smoothed minimum function is needed when fitting measured
behaviour in \cref{sec:benchmark}.
As the exact nature of the smoothing is not considered important,
different functions could be used.
In this work, the following function has been used, since it
only depends on a single parameter $k$:
\begin{equation}
  \label{eq:smin}
  \min{}_{k}(x,y) = -\frac{1}{k}\ln\left(e^{-k x} + e^{-k y}\right).
\end{equation}

\section{VME transfer time}
\label{sec:vme-transfer-time}

\begin{table}[!t]
  {\centering
  \caption{Transfer time of two different SBCs with various modules.}
  \label{tab:transfer}
  \begin{tabular}{clcccc}
    \toprule
    SBC & Module & SiCy & SiCy & BLT & MBLT \\
        &        &      & DMA  &     &      \\
    \midrule

    \parbox[t]{5mm}{\multirow{5}{*}{\rotatebox[origin=c]{90}{MVME}}}
        & DMA setup & - & \multicolumn{3}{|c|}{16.7 $(n = 1)$}\\
    \cmidrule{2-6}
        & MTDC-32 & 1.21 & & \\
        & MADC-32 & 1.26 & 0.49 & 0.44 & 0.33 \\
        & \vulom{} & \multirow{2}{*}{1.37} & \multirow{2}{*}{0.65} & \\
        & (\TRLOII{}) & \\
    \midrule
    \parbox[t]{5mm}{\multirow{7}{*}{\rotatebox[origin=c]{90}{RIO4}}} %
        & (M)BLT setup & - & - & \multicolumn{2}{|c|}{$6.5 + n \cdot 4.3$} \\
    \cmidrule{2-6}
        & MTDC-32 & 0.40 & & 0.17 & 0.09 \\
        & MADC-32 & 0.45 & & 0.22 & 0.12 \\
        & \vulom{} & \multirow{2}{*}{0.60} & &            \\
        & (\TRLOII{}) & \\
        & V785    & 0.50 & & 0.19 & 0.15 \\
        & V1190   & 0.45 & & 0.18 & 0.10 \\
    \bottomrule
  \end{tabular}\par
  \bigskip
  }
  \cref{tab:transfer} shows the measured transfer times in \si{\micro\s} per 32-bit word for two different
    SBCs and various modules.  The additional SBC setup overhead of
    starting DMA (direct memory access) or block transfers depend on how many, $n$, are
    scheduled at the same time.
    DMA allows multiple adjacent SiCy requests
    to be scheduled together on the MVME.
    The MTDC-32~\cite{MTDC} and MADC-32~\cite{MADC} are from Mesytec,
    the V785~\cite{caenV785} and V1190~\cite{caenV1190} from CAEN, and
    the \vulom{}~\cite{VULOM4b} from GSI and is running the \TRLOII{} firmware \cite{Johansson2013}.
\end{table}

The time it takes for the basic operation of transferring a \SI{32}{bit}-data word essentially determines $\tp$, $\too$, $\td$, and $\ts$.
Generally, VME access can be divided into two categories, namely single cycle (SiCy) and block
transfer, where the latter can transfer either 32 (BLT) or 64 (MBLT) bits at a time \cite{VITA1994}.
There is
also 2eVME and 2eSST \cite{VITA2003} (both are block transfer modes), but these are not widely supported in front-end modules.

SiCy transfers single data words with minimal overhead. Thus with module $i$ producing $d_{i}$ words, the data transfer time
with SiCy reads is
\begin{equation}
  \label{eq:SiCy-simple}
  \td = \sum_{i} d_{i} t_{i,\mathrm{SiCy}}.
\end{equation}
Since the designs of the individual modules also affect the word transfer times, $t_{\mathrm{SiCy}}$ has a module dependence, $t_{i,\mathrm{SiCy}}$.
While block transfers are asymptotically faster per word ($t_{i,\mathrm{block}} < t_{i,\mathrm{SiCy}}$), they may require significant setup times in the SBC:
\begin{equation}
  \label{eq:blt}
  \td = \sum_{i} \left(t_{\mathrm{setup}} + d_{i} t_{i,\mathrm{block}}\right).
\end{equation}
It is, however, possible to amortize this setup time by performing a so-called chained block
transfer, whereby data is read from multiple modules during the same transfer\footnote{Note
  that chained block transfers are not an additional VME transfer mode, but something some modules can be configured to arrange together, e.g.\, by using the IACK backplane chain for communicating a token between the modules.}:
\begin{equation}
  \label{eq:cblt}
  \td = t_{\mathrm{setup}} + \sum_{i} d_{i} t_{i,\mathrm{block}}.
\end{equation}
With $c_{i}$ being the number of reads required to determine the amount of data
and ensure synchronicity for each module,
$\too + \ts$ can be expressed analogous to \cref{eq:SiCy-simple}:
\begin{equation}
  \label{eq:ts}
  \too + \ts = \sum_{i} c_{i} t_{i,\mathrm{SiCy}}.
\end{equation}

\begin{figure}[!t]
  \centering
  \includegraphics[width=0.95\columnwidth]{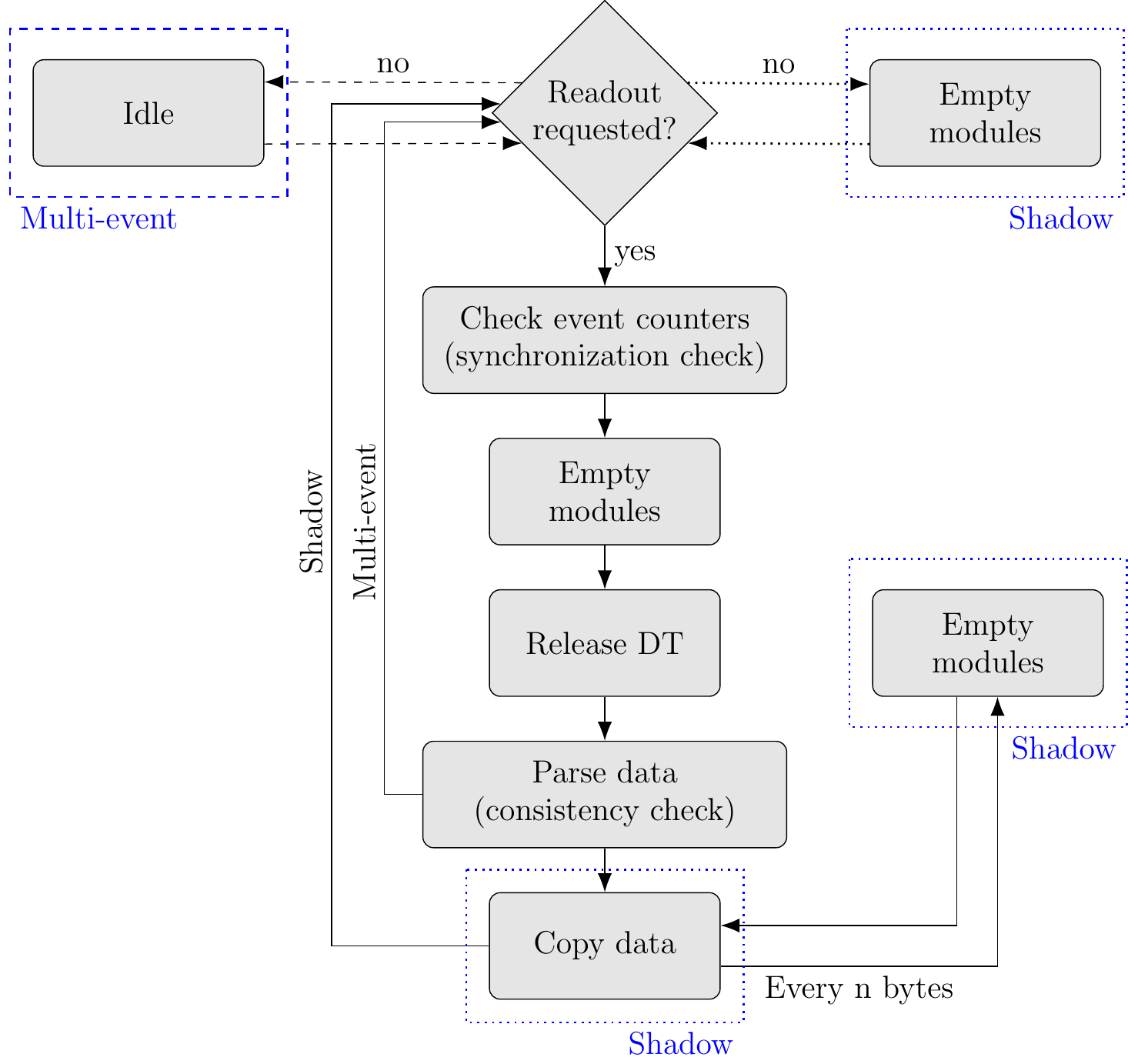}
  \caption{Flow diagram for multi-event and shadow readout.
    In multi-event mode, idle time is spent doing nothing or work for other processes.  In shadow mode, partial module readout is continuously performed.
  In both cases, the main readout will look
  for a readout request, which has asserted deadtime and will then transfer the (remaining) data and ensure that all
  modules are in sync. Then deadtime is released.
  Since data copy in shadow mode may be a lengthy process, the modules are regularly emptied during copying.
}
  \label{fig:flow}
\end{figure}

So far the discussion has been general. However, to evaluate actual systems, it is necessary to measure the
transfer times.
\Cref{tab:transfer} shows the measured read transfer time for a selection of
commercial modules when read using either
a Motorola\footnote{Motorola’s Embedded Communications Computing business
  was acquired by Emerson in 2007.}
MVME5500 \cite{mvmemanual} (MVME) or
a CES\footnote{CES was acquired by Mercury Systems, Inc. in 2016.}
RIO4-8072 \cite{rio4manual} (RIO4) SBC.
Both are operated with a Linux kernel \cite{mvmelinux, mbs62relnotes},
versions 2.4.21 and 2.6.33, respectively.
These transfer
times will be used in \cref{sec:benchmark} when modeling readout systems with these components.

Comparing the MVME and RIO4 SiCy transfer times in \cref{tab:transfer}, the RIO4 is generally three times faster per
word. Note that the timing differences between modules for the same SBC are roughly equal for the two
SBCs. This reflects the fact that it is the module VME access handling that determines these differences.

The measurements also clearly show that (M)BLT is significantly faster per word than SiCy, but has
a considerable setup overhead.

\section{Implementation}
\label{sec:concept}

In \cref{fig:flow} the work performed by the readout loop is sketched as a flow diagram.
Generally, the SBC will either poll a register or be notified via an interrupt that a readout
should be performed.  The trigger logic has then asserted deadtime.
When the readout request is found, the SBC will check all module event counters and
move the remaining data from all front-end modules to a CPU buffer. Then it can release the deadtime,
allowing the front-end modules to acquire more data while the SBC performs any further consistency checks
of the acquired data.
Afterward, it will resume waiting for the next readout request.

The difference between multi-event and shadow readout is the activities taking place
while waiting for a readout request. In the multi-event case, the SBC is mostly idle, except for network transfer tasks.

\begin{figure}[t]
  \centering
  \includegraphics[width=0.95\columnwidth]{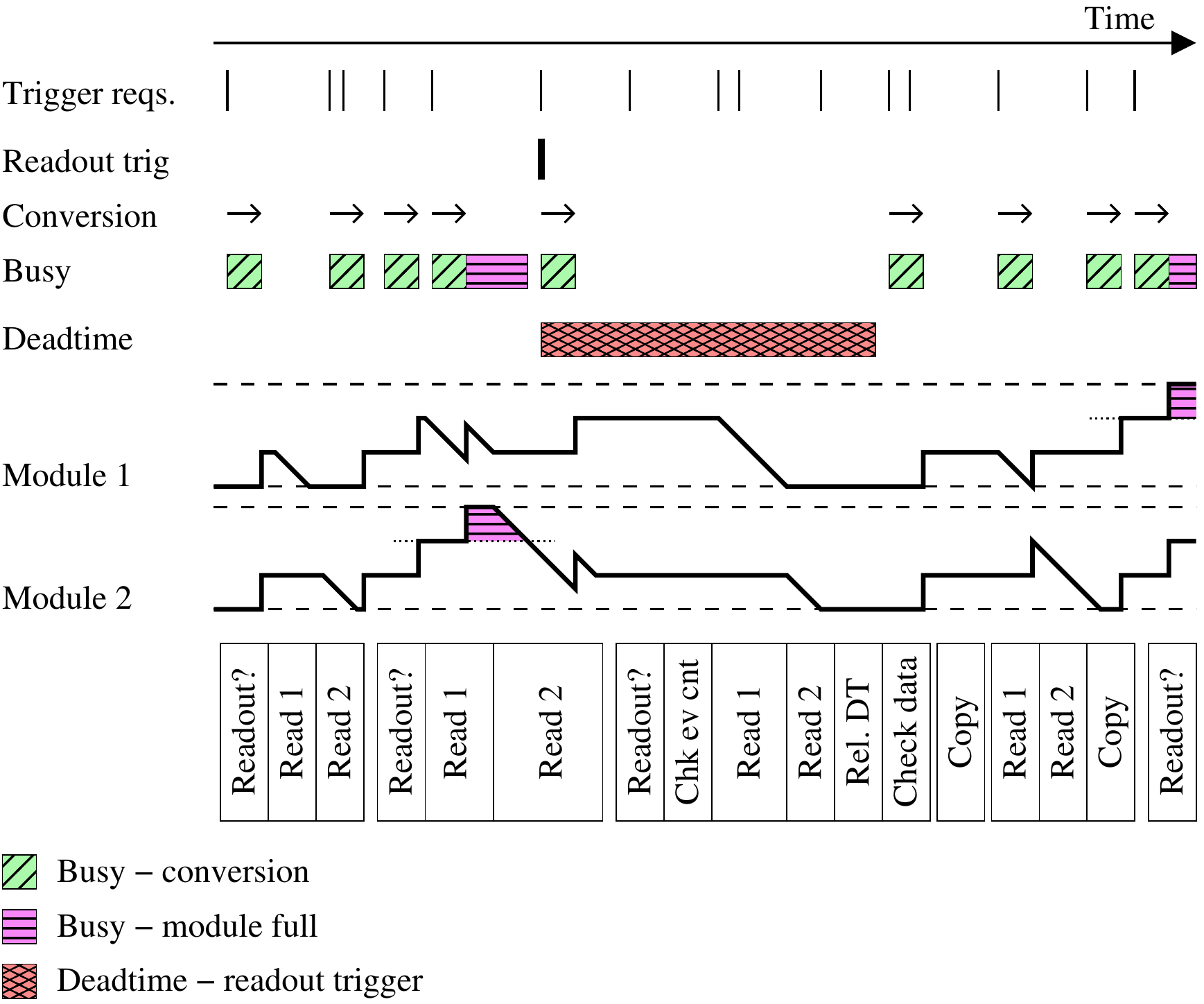}
  \caption{Example of SBC activity during shadow readout.
    The SBC repeatedly empties module buffers.
    Thus, when entering deadtime, the modules will be almost empty.
    While deadtime is asserted, event counters will be checked and
    the little remaining data transferred.
    After deadtime has been released, data will be consistency-checked and copied to the actual output buffer.
    The module graphs show their buffer fill levels.
    Busy is asserted both due to gate+conversion,
    and when any module buffer is full (above dotted lines).
    In this example, the module buffers can hold three events.
  }
  \label{fig:timeline}
\end{figure}

The shadow readout on the other hand tries to continually transfer data from the front-end
modules. This is illustrated for a two-module system in \cref{fig:timeline}, which also
shows the pausing when a module buffer becomes full and the deadtime management.

\begin{figure}[t]
  \centering
  \includegraphics[width=\columnwidth]{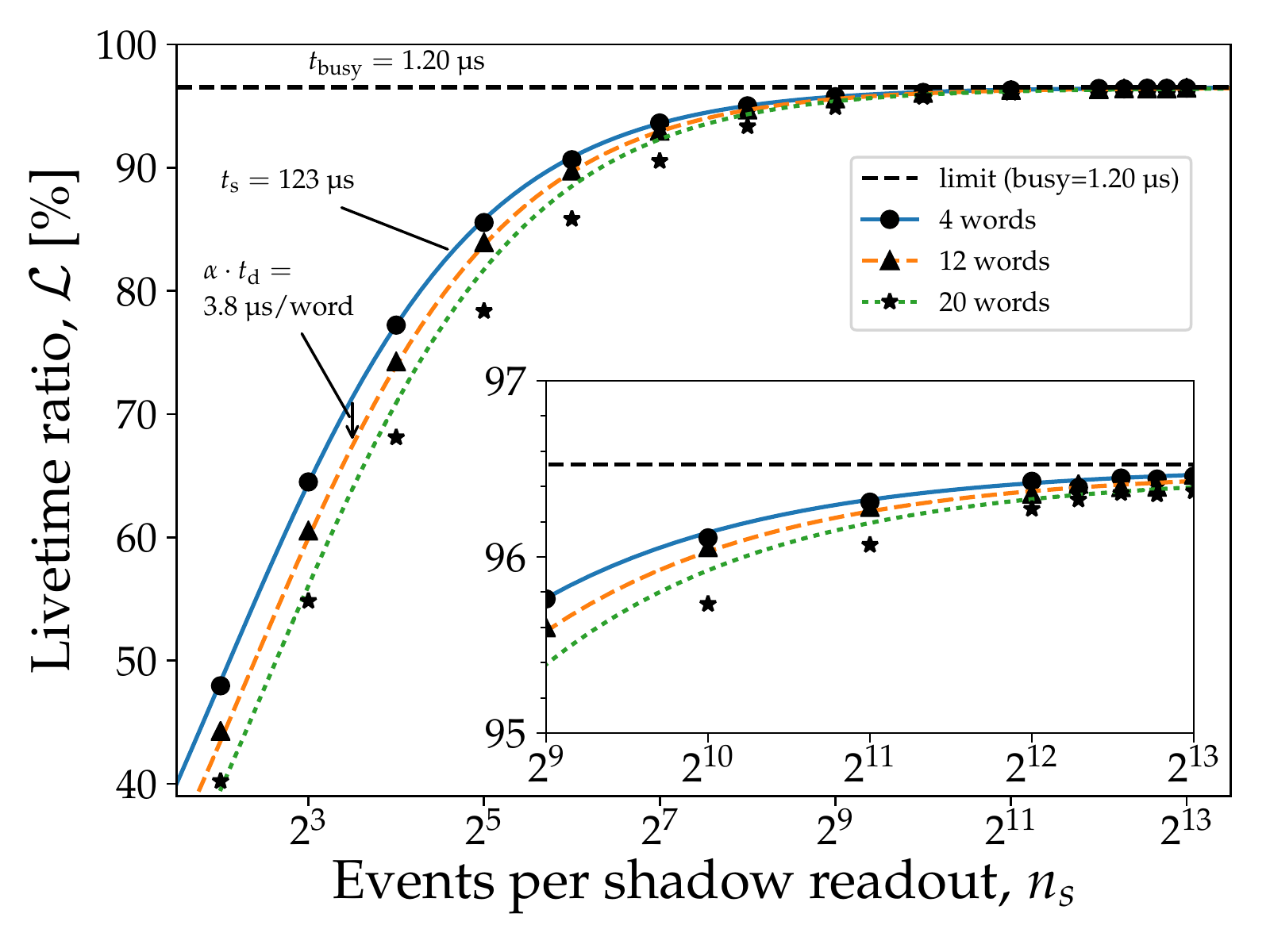}
  \caption{
    The \LT{} dependence on the number of events per shadow readout
    for an MVME and MADC-32 system with $\tg + \tc = \SI{1.2}{\us}$
    and $\fr = \SI{30}{\kHz}$.
    \LT{} converges to the conversion limit (dashed horizontal line)
    when $\ns$ is increased.
    With lower $\ns$, the overhead and synchronization time is amortized over
    fewer events.  With larger event sizes (varying $d_i$), the amount of remaining data to read
    during deadtime after each $\ns$ block of events increases.
    \revlabelnolineno{revmismatch20ns4}%
    The origin of the discrepancies for $d_i=\SI{20}{words}$ and $\ns \ge 2^4$
    is unknown.
  }
  \label{fig:shadow-scan}
\end{figure}

In the simplest case, the data from each module is transferred to a separate buffer.
This avoids potentially fragmenting events, and
the collected data can easily be consistency checked. 
In this fashion, it is no longer needed to limit $\ns$
to the module buffer depth $\nb$. Instead, the limitation for $\ns$ is the available RAM
and the desire to perform regular synchronicity checks.
The dependence on $\ns$ is illustrated in \cref{fig:shadow-scan},
\revlabel{revdatamodel}%
where measured data are shown with point-like markers,
and general trends of models are shown as curves. 
This scheme is used throughout the article.
The figure shows \LT{} for a system which has a conversion
time $\tc = \SI{1.2}{\us}$ (dashed horizontal line) and produced $d_i=4, 12,$ or $20$ words per event with
$\fr = \SI{30}{\kHz}$.
The figure shows
two essential features of the shadow readout: \LT{} converges towards the limit set by conversion
times when $\ns$ is increased. Additionally, \LT{} only shows a weak dependence on event size
with less than \SI{0.5}{\%} variation in \LT{} when $\ns = \SI{8}{k}$.

The virtual module buffers must eventually be copied to an output buffer. This is done after
the consistency check as shown in the flow diagram. However, since a substantial amount of data may
have been collected, this can easily take several milliseconds, which may be longer than module buffers
can continue to store new events,
causing them to assert busy signals. In order to avoid this, each
module is assigned two memory buffers operated in a ``ping-pong'' fashion such that new data is transferred into
the currently active buffer. When the modules have been emptied during deadtime, the other
buffer becomes active and will be filled with new data. The data from the inactive buffer is
copied in chunks interleaved with frequent calls to the shadow readout routine. This
is illustrated with the read of the modules in-between the copy blocks in
\cref{fig:timeline}.

We have implemented this readout scheme with a modified version of the readout library
nurdlib (formerly known as vmelib) \cite{Loher2014}.
When the readout polling is
performed by the surrounding DAQ framework code, it is necessary to also modify that. The 
change essentially amounts to one single line of code---a callback routine to allow data transfers in the background while no readout request is detected.

Hardware requirements are the same for multi-event and shadow readout, i.e.\ there must be an
acquisition control module that rejects events whenever deadtime or busy is asserted. This
module, which issues the readout requests, must be able to keep track of how many events have
been acquired and only request a deadtime-asserting readout after a sufficient number (\ns). In our
setup, all these tasks are performed by the
\TRLOII{} firmware
\cite{Johansson2013} running on the GSI \vulom{} module \cite{VULOM4b}.

To allow the most flexible use of the data from a multi-event readout system, it is
necessary also to be able to record for each trigger the detector
coincidences that caused the trigger to be selected.  This trigger
coincidence pattern is recorded for each event by the \TRLOII{} firmware.
In order to trust the system,
it is also necessary to verify that each trigger has
been seen by each module once (none lost, none spurious).  This is
done by comparing the event counters in the modules regularly, i.e.\ during each deadtime period.  This is enough since a correctly working
system would never have mismatches, making any deviation significant.
Nurdlib already performs these checks strictly.

Note that shadow readout causes almost continuous activity on the
VME backplane during analog conversion.  Within our tests,
it was no problem,
provided that the electronic modules and pre-amplifiers etc.\ were
properly grounded.  It is suggested to compare the noise levels
with and without shadow readout, e.g.\ using deadtime-asserting
multi-event readout for the latter.

\revlabel{revnoise}%

\begin{table}[tb]
  {\centering
  \caption{cpu time of network calls}
  \label{tab:networkoh}
  \begin{tabular}{lrl}
    \toprule
    SBC   & Buffer & Network CPU time\\
    \midrule
    \multirow{2}{*}{MVME5500}
          &                 32Mi & $\SI{1.4}{\us} + w \cdot \SI{8.8}{ns}$\\
          &                 64ki & $\SI{1.1}{\us} + w \cdot \SI{3.3}{ns}$\\
    \midrule
    \multirow{2}{*}{RIO4} & 32Mi & $\SI{4.5}{\us} + w \cdot \SI{9.5}{ns}$\\
          &                 64ki & $\SI{2.3}{\us} + w \cdot \SI{3.8}{ns}$\\
    \bottomrule
    E3-1286v6 &             32Mi & $\SI{0.41}{\us} + w \cdot \SI{0.091}{ns}$\\
    (Intel x86, 4.5 GHz) &  64ki & $\SI{0.37}{\us} + w \cdot \SI{0.065}{ns}$\\
    \midrule
  \end{tabular}\par
  \bigskip
  }
  \cref{tab:networkoh} shows the CPU time per network \texttt{write} call of $w$ bytes for
    two different SBCs, and a modern server CPU for reference.  The
    values are averages for either a large or a small buffer, the latter
    typically fitting in low-level CPU cache.
    The former are representative for DAQ network
    transfers from large event accumulation buffers.
\end{table}

\begin{figure}[!t]
  \centering
  \includegraphics[width=\columnwidth]{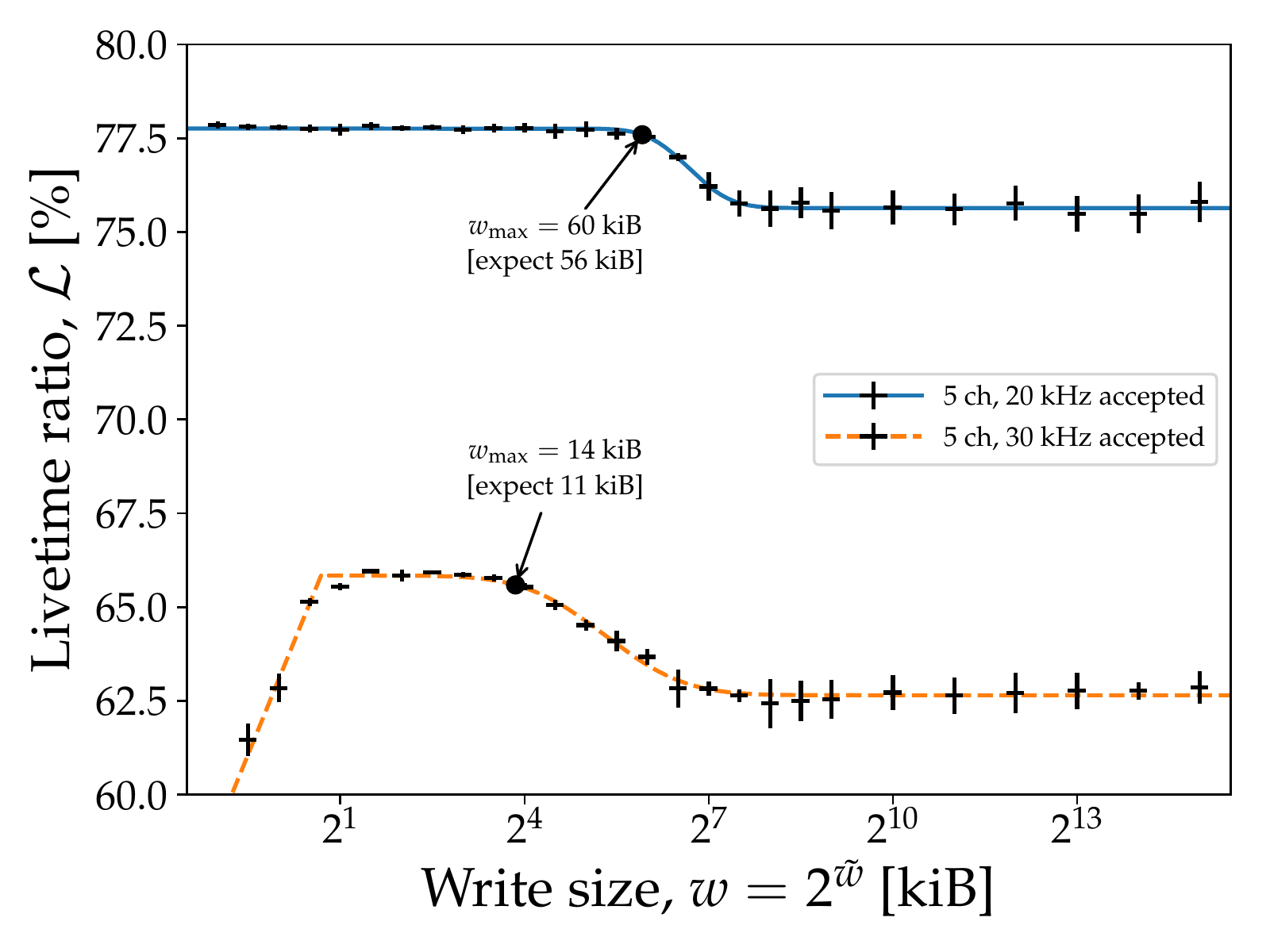}
  \caption{\LT{} as a function of network preparation chunk size $w$,
    for two different
    average trigger rates.  The system consists of a RIO4 with a \vulom{} and
    six CAEN V785s, delivering 33 words per event (5 channels per ADC).
    When the chunk size exceeds $w_\mathrm{max}$, the livetime ratio drops
    due to module buffers occasionally
    becoming full while the CPU is occupied with
    network processing.  Also, too small chunk sizes cause losses,
    due to CPU time lost on performing unnecessarily many calls.}
  \label{fig:writechunk}
\end{figure}

\section{Single core caveat}
\label{sec:single-core-caveat}

Both the MVME and RIO4 have single-core processors, which poses some
challenges related to the kernel scheduler of the operating system.
The main issue is the conflict between readout and network
transfer. The SBC must transfer the data to either an event builder or
non-volatile storage, meaning time not spent
emptying modules, see \cref{tab:networkoh}.
It will, therefore, lower the maximum rate that can be handled.  However,
before that, it can also impact the livetime of the system, if the
network preparation happens in such large uninterrupted chunks that
the module buffers become full.
This is shown in \cref{fig:writechunk}, where \LT{} drops by a few
percent in the naive case where the network transfer is issued
with too large buffers sent to the \texttt{write} system call.
The kernel takes too long to complete the request, which causes
the drop in \LT{}.
This can be mitigated by breaking the transfer preparation into
multiple chunks, each with a call to \texttt{write} directly followed
by yielding the timeslot of the network thread
(by \texttt{sched\_yield}).
This allows the readout thread to cycle through the modules for each
\texttt{write}.
The drop in \LT{} can be described by considering the maximum time each network preparation
chunk should take:
\begin{equation}
  t_\mathrm{n,max} = \beta \frac{n_\mathrm{b}}{\fmax} -
  \left(t_\mathrm{d} + \frac{t_\mathrm{o}}{n_\mathrm{b}}\right).
\end{equation}
The first term is the average time the module buffer can handle, and the second
term the CPU time spent on the readout.  The factor $\beta$ accounts for
the fact that a Poisson-distributed
trigger will sometimes issue a burst of many triggers in an unusually short time interval than the
long-term average, and thus fill the buffer more quickly.

The maximum time can, by using the values of \cref{tab:networkoh}, also be expressed
as a maximum chunk size, which can be used to control the network processing:
\begin{equation}
  w_{\mathrm{max}} =
  2^{\tilde{w}_{\mathrm{max}}} = \frac{t_\mathrm{n,max}}{t_\mathrm{write/B}}.
  \label{eqn:}
\end{equation}

The drop in the livetime ratio as $w$ exceeds $w_{\mathrm{max}}$ (as shown in
\cref{fig:writechunk}) can be fitted by
\begin{equation}
  \LT_1 = \LT_\mathrm{max} - \LT_\mathrm{drop}
  \left(\frac{1}{2}+\frac{1}{2}\mathrm{erf} \frac{\tilde{w} - \tilde{w}_{0}}{\tilde{w}_{\sigma}}\right).
  \label{eqn:writesize}
\end{equation}
We use the ``error function'' ($\mathrm{erf}$) to describe the slight but measurable drop in \LT{} around \texttt{write} size $2^{\tilde{w}_{0}}$.

\begin{figure}[!t]
  \centering
  \includegraphics[width=\columnwidth]{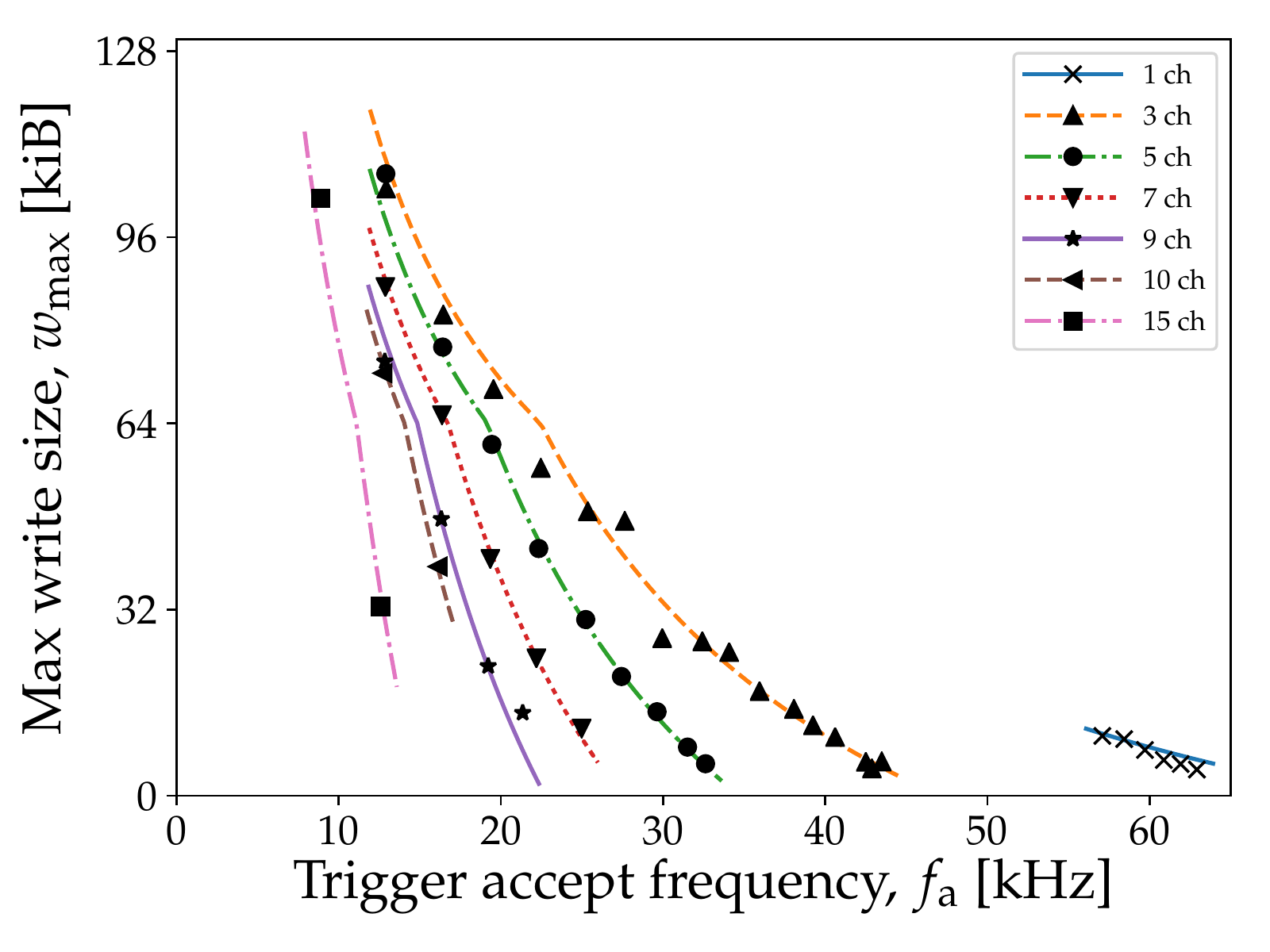}
  \caption{The maximum chunk write size $w_\mathrm{max}$ for the system
    in \cref{fig:writechunk}, determined for many combinations of event sizes
    and trigger rates.  The event sizes are varied by activating a different
    number of channels in the ADCs.}
  \label{fig:writechunkspred}
\end{figure}

\begin{figure*}[!t]
  \centering
  \includegraphics[width=\columnwidth]{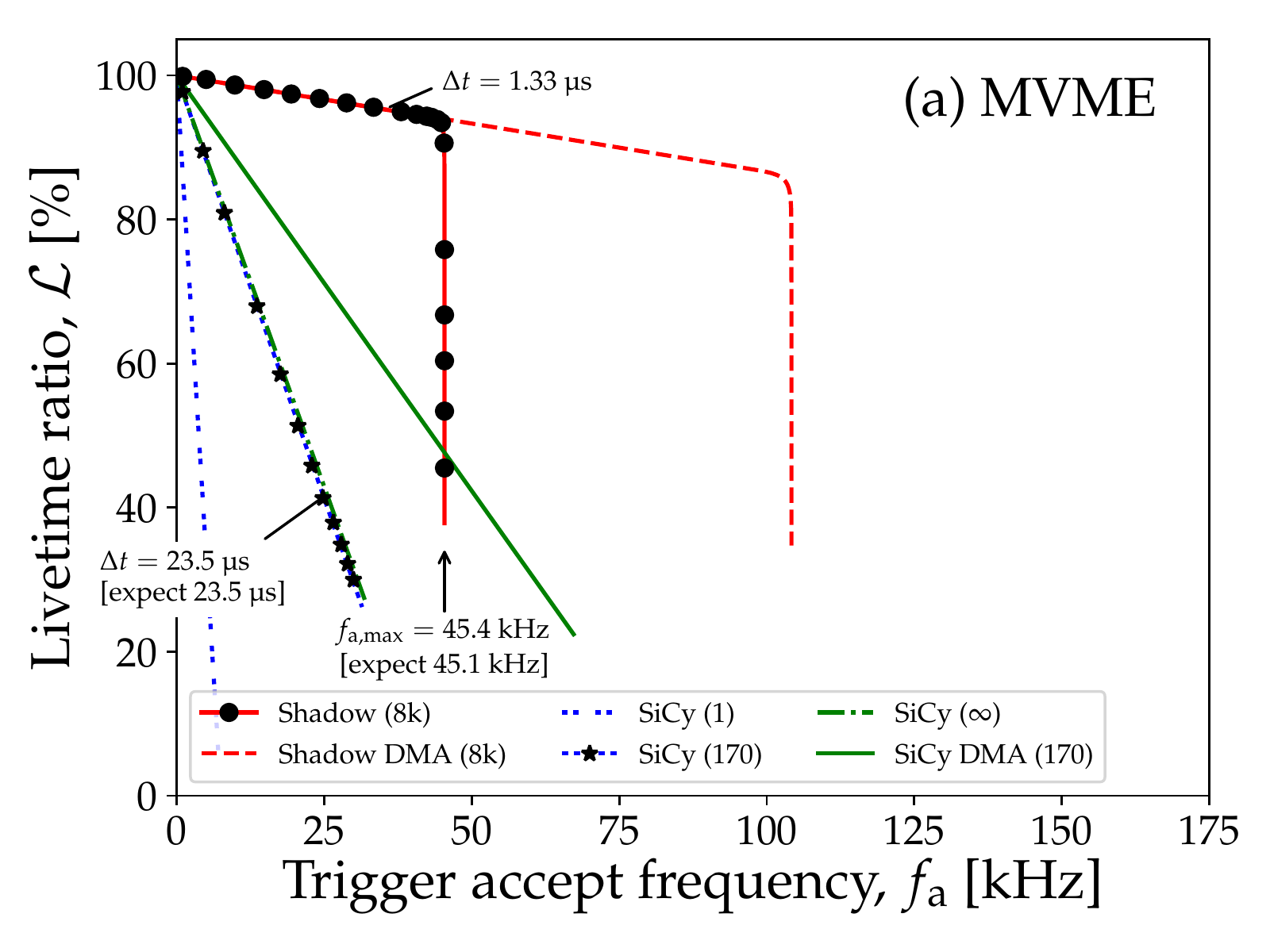}
  \includegraphics[width=\columnwidth]{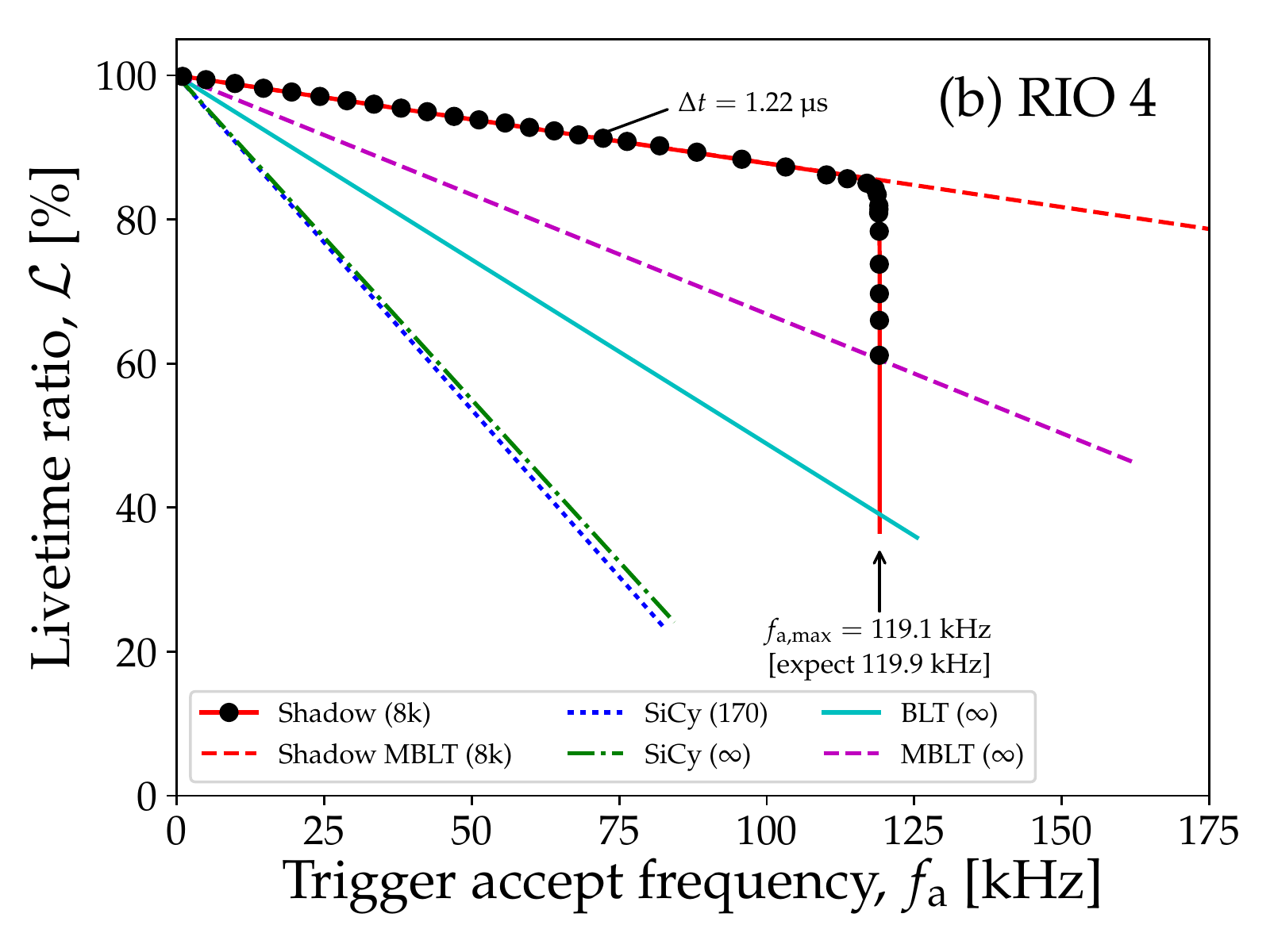}
  \caption{Livetime ratio \LT{} as a function of accepted trigger 
	frequency for the MVME and RIO4 SBCs
    reading from the \vulom{}, MADC-32, and MTDC-32 using
    various readout modes, with in total 17 words per event. The round data points show the measured \LT{} for shadow readout,
    while the stars are for multi-event mode, both with SiCy readout. The numbers in parentheses are
    the number of events accepted before the SBC readout function is invoked. $\infty$ corresponds
    to the limit of zero overhead.
    Note that the lower performance of the MVME SBC, see \cref{tab:transfer}, only affects the maximum number of events taken, \fmax, not the per-event deadtime at lower rates.
  }
  \label{fig:simple-bench}
\end{figure*}

Using too small chunk sizes, however, will introduce a steep %
loss, due to the overhead of very many \texttt{write} calls.
This is seen for the \SI{30}{\kHz} case in \cref{fig:writechunk}.

The above description has been checked for many
combinations of trigger frequencies
and event sizes, as presented in \cref{fig:writechunkspred}.
For each case, $w_{\mathrm{max}}$ has been determined from fits of
\cref{eqn:writesize} as in \cref{fig:writechunk} with the value
$2^{\tilde{w}_{\mathrm{max}}} = 2^{\tilde{w}_{0} - \tilde{w}_{\sigma}}$.
The curves thus describe estimates of the maximum \texttt{write} size
that can be used without affecting performance.
At buffer sizes above 64 kiB, the fit assumes that the network
overhead is twice as large, matching the measured data.  We have
however not been able to attribute this effect to a specific cause.
The best fit value of $\beta=0.79$ corresponds to bursts happening 
\SIrange{10}{1}{\percent} of times for $f_\mathrm{a}$ in the range
\SIrange{10}{50}{kHz}.
The percentages were obtained from simulations of Poisson-distributed
triggers taking a non-extending \SI{10}{\us} busy-time into account for stretches of $n_\mathrm{b}=32$ accepted triggers.

Even if the negative impact on the livetime is only a few percent,
mitigation is important since the effect
is enhanced for multi-node systems as the trigger requests
which become blocked on each node are independent.
This cumulative effect is illustrated in \cref{fig:caen-network}(b).

\revlabel{revsamebus}%
Provided that any common data bus between the CPU and the network and VME
interfaces can handle the simultaneous traffic,
this effect could be mitigated with a multi-core SBC.
One
core would be dedicated to the readout, while another would handle
non-critical tasks such as network transfers.

\section{System characterization}
\label{sec:benchmark}

In this section, we will benchmark the shadow readout system versus a regular multi-event
readout system using two different sets of modules. Each set is part of actual experiments. The
first system was used for the IS633 experiment at ISOLDE (CERN) \cite{Catherall2017} and the
second one is the current system used at the Aarhus \SI{5}{MV} accelerator. 

In all cases, the trigger requests are either provided by a CAEN DT5800D detector emulator
(pulser) \cite{UM3074}, which can provide a Poisson distributed trigger sequence with a given rate, or by equivalent functionality directly in the trigger logic firmware.
The trigger requests are sent to the \vulom{} running the \TRLOII{} firmware, which
handles the busy and deadtime logic.  It also provides scaler values for the total number of
trigger requests and the number of accepted requests.

Throughout the article, measured data are shown with point-like markers,
and general trends of models are shown as curves. For illustration, some models
are also evaluated for configurations that have not been
tested.

\subsection{IS633 system}
\label{sec:simple-system}

This is a simple configuration designed to run with very low deadtime. It consists only of an SBC,
\vulom, Mesytec MADC-32 and Mesytec MTDC-32. In order to achieve high livetime, the MADC is
configured in bank toggle mode \cite{MADC}. In this mode, the MADC toggles between which of its two ADCs
that digitize the signals and can thus accept a new event while still processing the previous one. The gate
was set to \SI{1}{\us} and the module in \SI{4}{k} mode which has a conversion time of
\SI{1.6}{\us}. The module will thus only assert busy if three events arrive within
a \SI{2.6}{\us} window. However, in practice, pile-up is best avoided, and thus \TRLOII{} emitted a
\SI{1}{\us} busy after every accepted event. 

The modules were configured such that the \vulom{} produced 3 words of data per event, the MADC-32 12, and
the MTDC-32 2. The busy was a logical OR between the three modules. The modules could
store 170, 481, and 2891 events, respectively. An additional 66 words (mainly scaler values) were produced and read once per
readout request. 

\subsubsection{Multi-event mode}
\Cref{fig:simple-bench} shows the livetime ratio %
as a function of the trigger request frequency when using either the MVME or RIO4. For the MVME,
\LT{} has been measured for multi-event mode using SiCy readout and a buffer depth of 170. The curve through the data points is \cref{eq:lt} with
$\Delta t = \SI{23.3}{\us}$.
This corresponds to an effective $\tc = \SI{1.3}{\us}$, the expected $\td = \SI{21.5}{\us}$, and a combined overhead
per event of $\SI{0.53}{\us}$. The green dash-dotted curve shows the expected behavior with zero
overhead ($\infty$ multi-events) and it would only provide a slight improvement since the data transfer
dominates the deadtime (curve overlaps with SiCy(170) in the figure). The RIO4 system would %
perform significantly
better with a SiCy readout time per event of only \SI{8}{\us}. \Cref{fig:simple-bench}(b) also
shows the curves corresponding to 32- and 64-bit block transfer in the limit of zero
overhead per event. Using block transfer, the per-event readout time would be reduced to 4.1 or
\SI{2.3}{\us}, respectively.

\subsubsection{Shadow mode}
\Cref{fig:simple-bench} also shows the measured \LT{} for shadow readout doing SiCy reads with a shadow buffer
depth of 8192 events.
Up to \fmax, the data points follow \cref{eq:lt} with $\Delta t$
roughly equal to the gate time.
According to the models discussed earlier (\cref{eq:fmax} %
together with values from \cref{sec:vme-transfer-time}), the maximum accept frequency
the MVME can sustain is
\SI{46.4}{\kHz} while the RIO can handle \SI{125}{\kHz}.
Note that up to the respective \fmax, the MVME and RIO4 deliver essentially the
same livetime performance.

If the maximum bandwidth of SiCy transfers becomes the limiting factor, one could combine block transfer with
shadow readout. The red dashed curves labeled \emph{Shadow DMA} or \emph{MBLT} in \cref{fig:simple-bench} show such
configurations. In this case, the limiting factor is the block transfer overhead, with an
expected $\fmax = \SI{388}{\kHz}$ for the RIO4 with this particular setup and number of channels.

\subsection{Aarhus system}
\label{sec:aarhus-system}

The other system consists of a RIO4, \vulom, and six CAEN V785 ADCs. This system is used at the
Aarhus University \SI{5}{MV} accelerator.  To mimic actual production conditions, each ADC module was
configured to produce 5 words per event while the \vulom{}
produced 3. In total 33 words were produced per event with an additional 66 words produced once
per readout. 

The CAEN V785 has a 32-entry multi-event buffer 
and a total conversion time of \SI{7.06}{\us}, which includes settling times etc. Additionally,
a \SI{3}{\us} gate is used. Hence, the expected deadtime per trigger is
\SI{10.06}{\us}. It should also be noted that chained block transfer is not
advantageous for the V785, as it only can deliver one event per transfer for these modules. Since we have more
events than modules, we will instead model the simultaneous scheduling of 6 block
transfers. This has an overhead of \SI{32}{\us}, i.e.\ \SI{1}{\us} per event. 

\begin{figure}[t]
  \centering
  \includegraphics[width=\columnwidth]{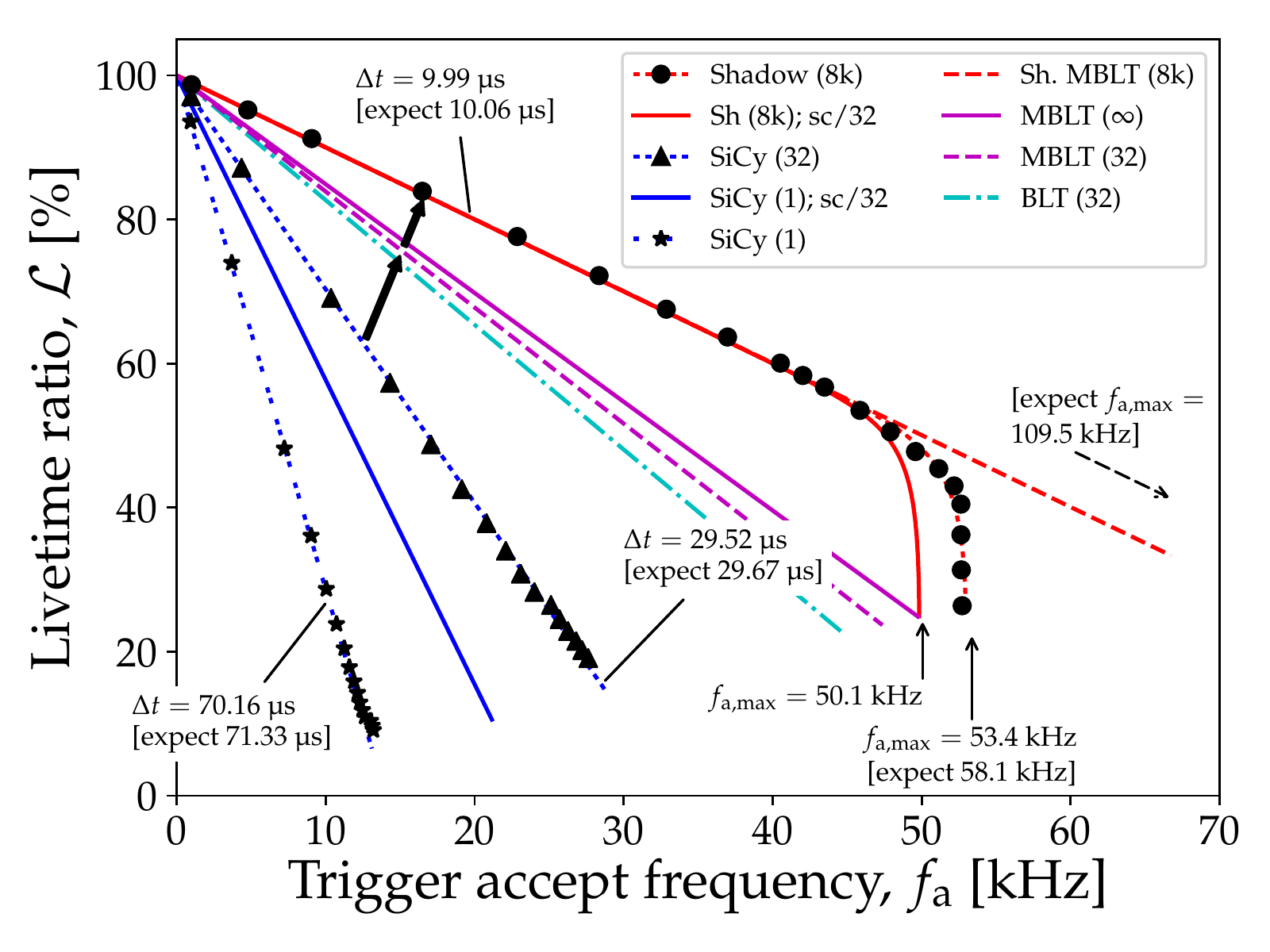}
  \caption{Livetime ratio \LT{} as a function of accepted trigger 
	frequency for a RIO4 with a \vulom and
    six CAEN V785s for various readout modes.  Each event consists of 33 words. The round data points show the measured
	\LT{} for shadow readout. The numbers in parentheses are the 
	number of events accepted before
    the SBC readout function is invoked. $\infty$ corresponds to the limit of zero overhead.  sc/32 denotes that scaler data is recorded every 32 events, providing a fair comparison with multi-event readout.
    The values displayed are from fits, while the expected values are predictions based on the values in \cref{tab:transfer}.
    The thick arrows indicate the improvement obtained over the previously
    used multi-event mode at working conditions
    of \SI{20}{kHz} requested trigger rate. The deadtime ratio is halved
    and the accepted and thus recorded trigger rate
    increased by \SI{30}{\percent}.}
  \label{fig:caen-bench}
\end{figure}

In practice, readout with (M)BLT of the CAEN V785 is %
troublesome with some modern SBCs, since
the data length cannot be queried beforehand and instead is marked with a VME \texttt{BERR*}
signal.  While allowed by the standard, this interacts badly with some SBC block transfer drivers since they do not report the number of actually transferred words, and thus obliterates much of the
benefits.
In the model, we have assumed that
the SBC driver provides the needed information.

Another issue with (M)BLT of the CAEN V785 is that the busy release
after conversion is delayed until the ongoing VME transfer has
completed  \cite{caen_priv_apr2018}.
This will inflate the effective conversion time.  This
effect has also been ignored in the model.

\Cref{fig:caen-bench} shows the results obtained for the shadow readout and the measured and estimated
behavior of various multi-event modes. The SiCy data transfer time per event is \SI{17}{\us},
while BLT and MBLT take \SI{6.3}{} and \SI{5}{\us}, respectively. This includes the block transfer
overhead of $\sim15\%$
compared to the transfer time. 
However, even in the limit of zero overhead, block transfer does not converge to the limit of
conversion and gate time. On the other hand, the data obtained show that SiCy shadow readout
can maintain the limit and keep up with the data rate until $\sim \SI{53}{\kHz}$, and
to $\sim \SI{50}{\kHz}$ when scaler readout for every 32nd event is included.

\subsection{Network impact}

\begin{figure}[t]
  \centering
  \includegraphics[width=\columnwidth]{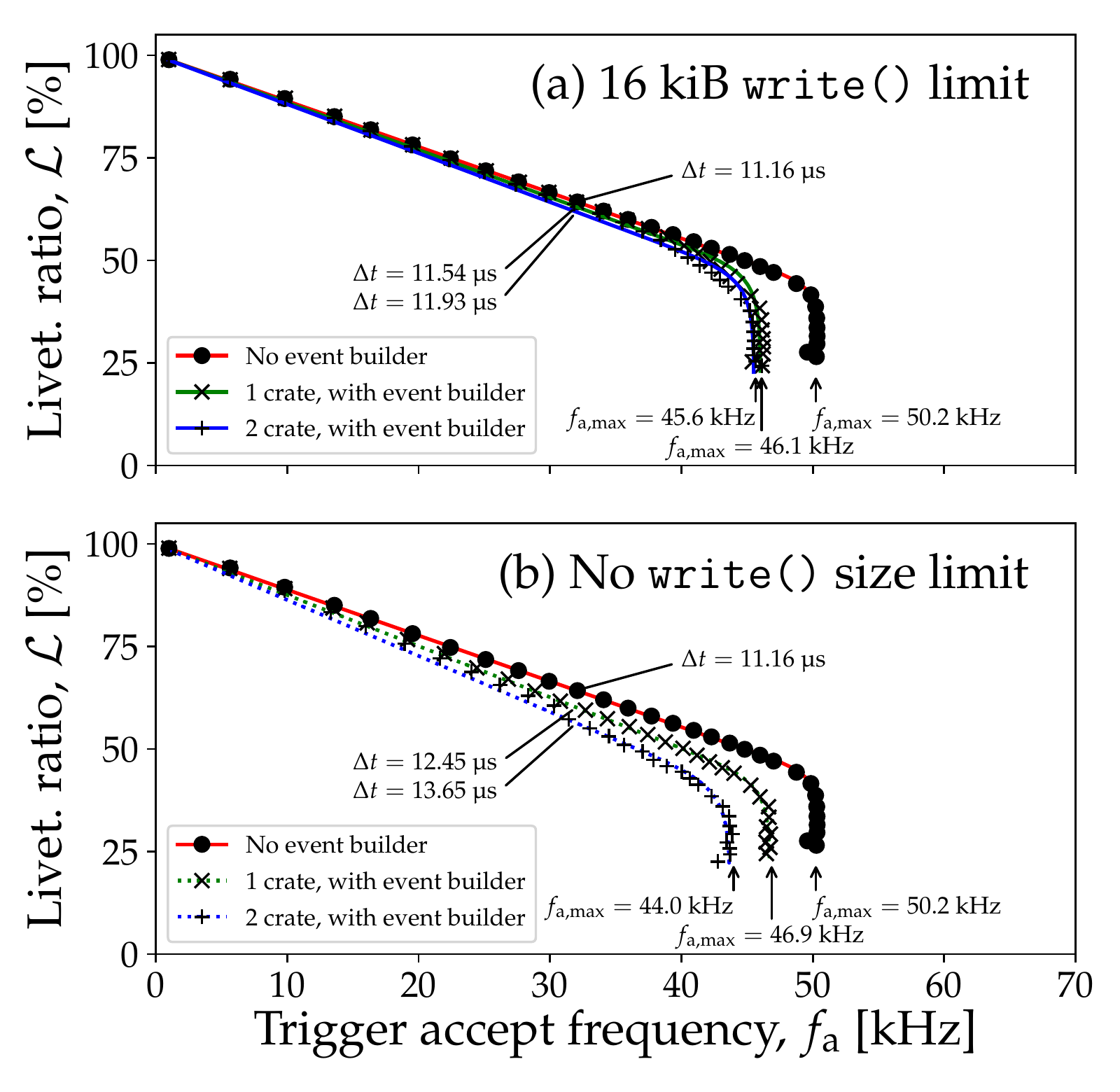}
  \caption{Livetime ratio \LT{} as a function of accepted trigger 
	frequency for
    the same system as in \cref{fig:caen-bench}, with network
    transport of data to a common \revlabelnolineno{revEBcaption}event builder
    (cross markers), and
    a multi-crate system with two duplicate crates
    (plus markers).
    The measurement
    without network transport (circles) is also shown.
    In (a), the network \texttt{write} calls are limited
    to 16 kiB.
    Note that before \fmax{} is reached, \LT{}
    is in all cases essentially given by \BT{}, i.e.\ it is virtually unaffected by the network
    transport.
    In (b), the network \texttt{write} calls have no limit
    and thus cause additional deadtime due to module buffers becoming
    full while the CPU is busy with network processing.
    The additional deadtime
    occurs for independent events, accumulating the effect for
    multi-crate systems.}
  \label{fig:caen-network}
\end{figure}

The above measurements were done without network transport, in
order to simplify the system descriptions.  The data rates, at a few
MB/s, do not use any significant CPU resources for network processing.
This is seen in \cref{fig:caen-network}(a)
for a single-crate system with and without network transport,
where the network overhead due to sending $\SI{6.3}{MB/s}$ in
$\SI{16}{kiB}$ chunks should use $6.1\%$ CPU time according to \cref{tab:networkoh}.  The
lowering of $\fmax$ from $\SI{50.2}{kHz}$ to $\SI{46.1}{kHz}$ corresponds
to $\SI{8.2}{\percent}$.

\subsection{Multi-crate system}

The shadow readout mode can also be applied to multi-crate systems.
The Aarhus system described above (RIO4, \vulom, and six CAEN ADCs)
was duplicated
in a second crate, and is operated together with the first in a master-slave configuration.
The data from both systems are sent to an event builder PC for merging.
The master start signal is used to generate gates for the modules
in both crates for each trigger.
The master crate generates the readout triggers, and by means of a
TRIVA \cite{TRIVA} mimic connection \cite{Johansson2013} they are distributed to be handled simultaneously by the slave
system.  The readout deadtime is the logical OR combination
of the deadtime in the
two systems.  The busy signals are handled similarly, i.e.\ busy
reported by any module in the total system inhibits further
triggers.

The upper plot in \cref{fig:caen-network} shows that the trigger
and shadow readout operation is virtually unaffected by the second
crate provided that the execution time of each network \texttt{write}
call is limited.  The lower plot shows how stalls due to too long network
processing that happen at independent events cause losses
that scale with the number of involved readout nodes.

\subsection{Multi-crate event correlations}

\begin{figure}[!t]
  \centering
  \includegraphics[width=\columnwidth,trim={0.1cm 0.7cm 0.1cm 0.95cm},clip]{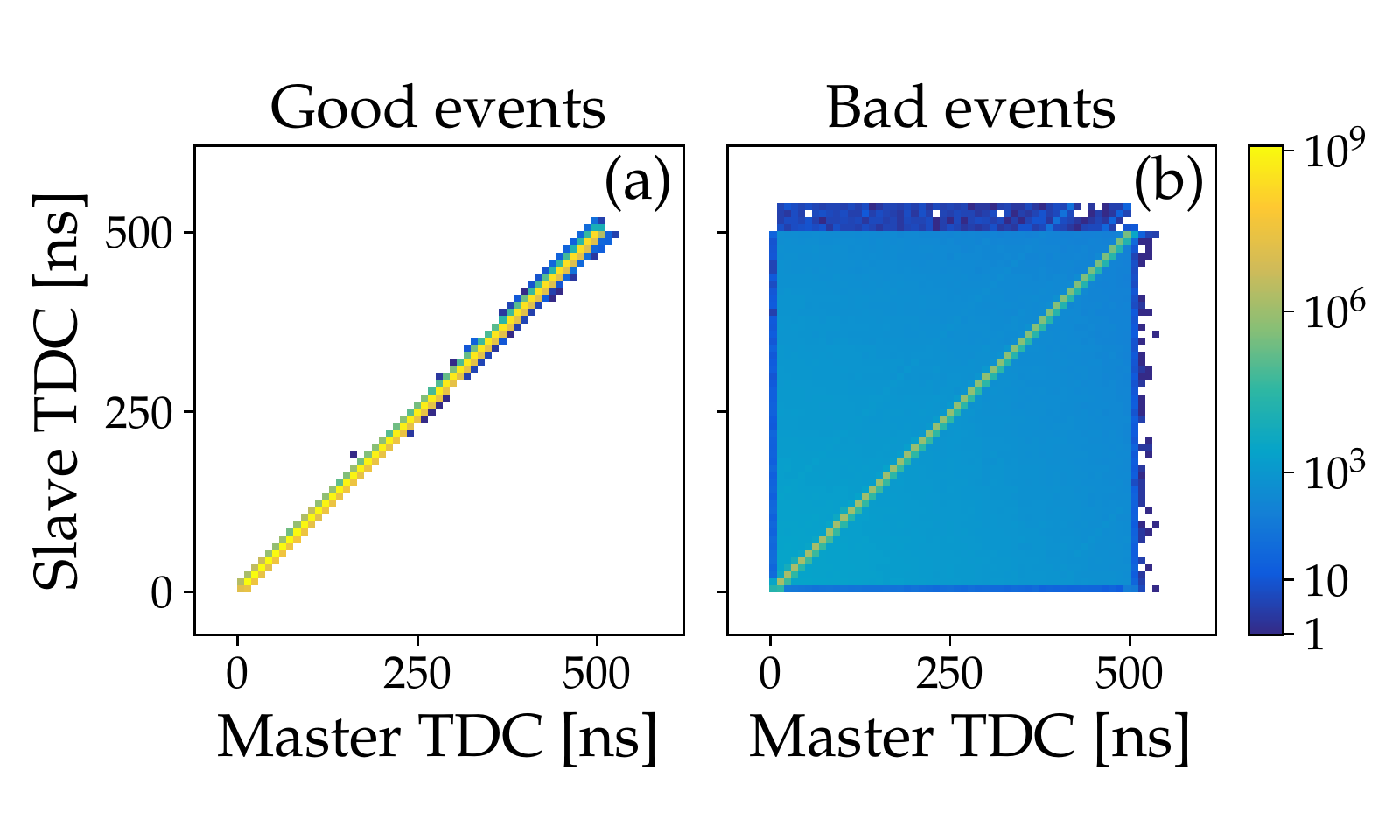}
  \caption{
    \revlabelnolineno{revsubplotlabels}%
    The clean diagonal line demonstrates the ability to
    detect and discard spurious triggers in a multi-crate system.
    Shown is the simultaneous measurement of $3.1\cdot10^{10}$ events with the same signal
    in both the master and slave system.
    Here, the slave system has
    in addition to the correct common triggers also intentionally
    received a low rate of spurious, i.e.\ wrong, master starts.
    This gives rise to $4.2\cdot10^{7}$ bad events.
    Potentially bad events are,
    within each shadow readout block of events,
    detected by event count mismatches
    between the data from the systems.
  }
  \label{fig:correlation}
\end{figure}

In any multi-event readout scheme, event mixing is a potential 
cause of severe data corruption, which can look deceptively correct.
By having taken multi-event to the extreme, shadow readout is
particularly exposed to this.
Multi-crate systems naturally also have more vulnerable components.

To show that the regular synchronisation and event counter checks are
effective against unintentional event-mixing due to spurious triggers
also in shadow readout multi-crate systems,
a two-crate system, each with one TDC (CAEN V775), was set up.
The TDCs are as usual given the same start signal, and one channel in each TDC
is fed the same stop signal, which should give a 1:1 correlation graph for
corresponding events, see \cref{fig:correlation}(a).
The stop signal is generated from the start
with a random delay of \SIrange{0}{500}{ns}.

To inject errors,
spurious triggers are injected with about \SI{1}{Hz} in the slave
system and its TDC.
This introduces additional events from the slave TDC, which
cause following events to be erroneously combined between the two
systems, until the next synchronization check.

Potentially bad events are
separated from the presumed good ones during analysis by
the detection of event counter mismatches between the master and slave
systems.
This is seen in \cref{fig:correlation}(a), where no mixed
correlations are observed among the presumed good events.
The bad events still have a pronounced diagonal, since only events in a
readout block after the spurious trigger are affected, while the
detection granularity is entire readout blocks.

\section{Conclusion and outlook}
\label{sec:conclusion-outlook}

Three principle ways to operate VME-based readout systems for multi-event capable digital acquisition modules
have been described. These schemes are single-event, multi-event, and shadow readout. The
necessary considerations to describe the performance of these schemes in terms of total system deadtime, based on the timing of individual operations, have been
detailed. From these considerations, it is expected that the deadtime ratio for a scheme in
which data readout is performed asynchronously to the conversion should converge to the limit of the busy
time of the front-end electronics. An implementation of such a scheme was then presented in some detail, and it was shown that
the deadtime ratio {\revlabel{revhyphen}}converges as long as the VME bus has sufficient bandwidth for the total data
rate, using either single-cycle access or block transfer. Finally, a benchmark of two different systems with shadow readout was shown.  In both
cases, the shadow readout scheme achieved higher livetime than the alternative readout schemes, even when the latter used faster
block transfer modes.

At the time of writing, a shadow readout system has been running for several months at the Aarhus University \SI{5}{MV} accelerator without problems, and the methods have been successfully used for the IS633 and IS616 experiments at
CERN ISOLDE.

\section*{Acknowledgment}
\label{sec:Acknowledgment}
\addcontentsline{toc}{section}{Acknowledgment}

The authors would like to extend their thanks to Dr. N. Kurz for the
work to provide a stable Linux environment with block transfer modes
for the RIO4 SBC.
The authors would like to extend their thanks to the Daresbury NPG
for providing the MVME Linux environment.

\bibliography{IEEEabrv,shadow-article}

\end{document}